\documentstyle[
12pt]{article}
\newcommand{\simlt}  {\raisebox{-.6ex}{$\stackrel{\textstyle <}{\sim}$}}
\newcommand{\simgt}  {\raisebox{-.6ex}{$\stackrel{\textstyle >}{\sim}$}}
\begin{document}
\begin{flushright}
RAL-94-072  \\
13 June 1994 \\
\end{flushright}
\vspace{5 mm}
\begin{center}
{\Large Generation Permutation Symmetry and the Quark Mixing Matrix}
\end{center}
\vspace{5mm}
\begin{center}
{P. F. Harrison \\
Physics Department, Queen Mary and Westfield College\\
Mile End Rd. London E1 4NS. UK \footnotemark[1]}
\end{center}
\vspace{1mm}
\begin{center}
{and}
\end{center}
\vspace{1mm}
\begin{center}
{W. G. Scott \\
Rutherford Appleton Laboratory\\
Chilton, Didcot, Oxon OX11 0QX. UK \footnotemark[2]}
\end{center}
\vspace{1mm}
\begin{abstract}
\baselineskip 0.6cm
We propose a new discrete symmetry in the
generation space of the fundamental fermions,
consistent with the observed fermion mass spectrum.
In the case of the quarks, the symmetry leads
to the unique prediction of a flat CKM matrix
at high energy.
We explore the possibility that evolution due to
quantum corrections leads to the observed
hierarchical form of the CKM matrix at low energies.
\end{abstract}
\begin{center}
{\em To be published in Physics Letters B.}
\end{center}
\footnotetext[1]{E-mail:PFH@V1.PH.QMW.AC.UK}
\footnotetext[2]{E-mail:SCOTTW@RL.AC.UK}
\newpage
\baselineskip 0.6cm
The problem of the origin of
the masses and the mixing angles
of the fundamental fermions must surely be amongst
the most urgent in particle physics today.
Even accepting the standard mechanism for
fermion mass generation through Yukawa couplings
to one or more non-zero Higgs fields, the reason
for the existence of three fermion generations
together with the explanation for the observed
pattern of the individual masses and mixing angles
remains mysterious.
One possible way forward is to gain
experience by constructing and analysing
a wide variety of plausibly motivated candidate mass matrices
(or ansatze) in the hope that something
convincing will eventually emerge.
Amongst the best known and perhaps the most thoroughly
analysed such ansatz is that due to Fritzsch \cite{FRITZCH}.
The present proposal
has more in common with the approach pioneered by
Harari et al.\ \cite{HARARI}.

In this paper we motivate and analyse
a new ansatz for the fermion mass matrices, which
we believe has unique a priori appeal
by virtue of the principles underlying its construction.
Our proposal owes something to the straightforward
and oft-repeated observation that the fermion generations
are in some (yet to be defined) sense
duplicate copies one of the other.
That is to say that,
in spite of the large mass differences
observed from generation to generation,
it is natural to assume that
the three generations exist
fundamentally on an equal footing.
In constructing our ansatz,
we take this notion seriously
and insist that, at the most fundamental level,
there be no physical basis for prefering one
generation over another,
ie.\ in the Lagrangian the assignment of the
generation labels ($i=1$-3) must be entirely arbitrary.
Such a demanding requirement has much
of the character of established invariance principles in physics,
and naturally puts very severe constraints on the
form that the mass matrices can take.
Indeed these constraints
are so severe that they can often appear at first sight
to be in conflict with the experimental facts.
We show in this paper, however,
that this is not neccessarily the case.
The indisputable a priori appeal of the above idea,
taken together with the uniqueness and economy
of its implementation, have provided
much of the motivation to pursue this analysis.

We begin by noting that a principle of the sort outlined above
is trivially satisfied by the charged-current
weak interaction in any weak basis, as a consequence
of the universality of the weak interaction.
On the other hand, the evident large mass differences
observed, from generation to generation, tell us that
the Yukawa couplings in the physical basis,
are quite definitely not universal.
At this point,
the {\em only} solution that we can see,
consistent with the principle we have expounded above,
requires that we postulate that in some weak basis
the Yukawa couplings for a given fermion species
exhibit an invariance under {\em permutations}
of the generation indices.
A candidate mass matrix fulfilling our
requirement, which is also hermitian is:
\begin{equation}
m=\left(\matrix{
a & b & b^{*} \cr
b^{*} & a & b \cr
b & b^{*} & a \cr
} \right)
\end{equation}
where $a$ is real and $b$ is complex.
Note that the diagonal mass terms are all identical
(they are all equal to $a$) and that the off-diagonal
(weak-generation-changing) amplitudes for the
`clockwise' transisitions ($1 \! \rightarrow \! 2$,
$ \! 2 \rightarrow \! 3$ and $3 \! \rightarrow \! 1$)
are also all identical (they are all equal to $b$)
and the amplitudes for the
`anticlockwise' transistions ($ 1 \! \rightarrow \! 3$,
$3 \! \rightarrow \! 2$, and $ 2 \! \rightarrow \! 1$)
are all equal to $b^*$, so that no generation is preferred.
A matrix of this form is sometimes referred
to as a circulant \cite{MATHS}.
It might be argued that the mass matrices are unlikely
to be hermitian and that a general circulant matrix with
$a$ complex and with unrelated complex numbers $b$ and $c$
representing different amplitudes for the clockwise and anticlockwise
transisitions, would also satisfy our requirement.
Nothing is to be gained, however, by postulating this
general form since, on taking
the hermitian square ($mm^{\dagger}$),
we immediately recover the form eq.(1),
and, as is well known,
only the hermitian square of the mass matrix
can influence the measured masses and mixing angles.

Suppose that we postulate a matrix of the above
form for the hermitian square of the mass matrix
for the charged leptons.
The observed mass spectrum can be reproduced by setting:
\begin{eqnarray}
a&=&(\tau /3) +(\mu /3) + (e/3) \nonumber \\
b&=&(\tau /3)\ \omega_1 + (\mu /3)\ \omega_2 + (e/3)\ \omega_3
\end{eqnarray}
where $\tau$, $\mu$ and $e$ represent the masses-squared
of the $\tau$-lepton, muon and electron respectively,
and the $\omega_i$, $i=$ 1-3
are the usual complex cube-roots of unity.
In this form, in the rank-1 limit
($\mu,e \rightarrow 0$) the above matrix
reproduces the matrix proposed by Harari et al.\ \cite{HARARI}.
The form of eq.(2) follows from the general result
that the spectrum of the
eigenvalues of a circulant matrix is
given by the (discrete) Fourier transform
of its trailing diagonal.
The eigenvectors of a matrix of the form eq.(1) are:
(1,1,1), (1,$\omega_2$,$\omega_3$), (1,$\omega_3$,$\omega_2$).
These are of course just the momentum eigenstates
for a three-point one-dimensional lattice
satisfying periodic boundary conditions.
An operator of the form eq.(1) (with $b$ real and negative)
was employed by Feynman \cite{FEYN}
to describe the low lying energy states of the
tri-phenyl-cyclo-propanyl ion.
We consider it very significant that
the matrix operator defined by eq.(1) and eq.(2)
has so much in common with the simple derivative operators
representing the ordinary kinetic terms in the Lagrangian,
which as a consequence of translational invariance may
also be represented by (infinite) circulant matrices.
It might also be worth noting that the form eq.(1)
may equivalently be regarded as the $3 \times 3$ generalisation
of the phenomenologically successful $2 \times 2$ effective-theory
\cite{PAIS} used to describe the properties of the neutral kaon system,
prior to the discovery of CP violation.

Turning now to the quark mass matrices
one might be tempted to postulate mass matrices of the form eq.(1),
but with different parameters $a$ and $b$, chosen in analogy
with the case of the leptons above, so as to reproduce
the observed mass spectrum for the up-type and down-type
quarks respectively.
But matrices of the form eq.(1) commute with each other
for all values of $a$ and $b$,
so that the mass matrices for the up-type and down-type quarks
would be simultaneously diagonalisable and
the quark mixing (CKM \cite{CKM}) matrix would then be
the identity (or a trivial permutation matrix),
in clear disagreement with experiment.

With these considerations in mind, we have
investigated mass matrices of the somewhat more general form:
\begin{equation}
m=\left(\matrix{
a & be^{i\phi_3} & b^{*}e^{-i\phi_2} \cr
b^{*}e^{-i\phi_3} & a & be^{i\phi_1} \cr
be^{i\phi_2} & b^{*}e^{-i\phi_1} & a \cr
} \right)
\end{equation}
with $a$ and $b$ still given by eq.(2) and
with $\phi_1+\phi_2+\phi_3 = 0$,
so that the mass eigenvalues are unchanged.
In eq.(3) the off-diagonal amplitudes
are equal in magnitude but differ in phase,
so that the matrix eq.(3)
does not commute with the matrix eq.(1),
nor does it commute with matrices of the form eq.(3)
with different values for the phases.
The eigenvectors of a matrix of the form eq.(3) are:
($1,e^{-i\phi_3},e^{i\phi_2}$),
($1,\omega_2e^{-i\phi_3},\omega_3e^{i\phi_2}$),
($1,\omega_3e^{-i\phi_3},\omega_2e^{i\phi_2}$).
If we postulate matrices of the form eq.(3)
for (the hermitian squares of) the mass matrices
for the up-type and down-type quarks, and
construct unitary matrices $U$ and $D$
comprising the respective mass-ordered normalised eigenvectors,
we find that the CKM matrix ($V=U^{\dagger}D$)
may then itself be written as a circulant:
\begin{equation}
V=\left(\matrix{
p & q & r \cr
r & p & q \cr
q & r & p \cr
} \right) .
\end{equation}
Observables depend only on the phase
differences ($\Delta \phi_i$) between the corresponding
amplitudes in the up-type and down-type mass matrices:
\begin{eqnarray}
|p|^2 & = & (3+2{\rm Re}S)/9 \nonumber \\
|q|^2 & = & (3-{\rm Re}S+\sqrt{3}{\rm Im}S)/9 \\
|r|^2 & = & (3-{\rm Re}S-\sqrt{3}{\rm Im}S)/9 \nonumber
\end{eqnarray}
with $S=e^{i\Delta \phi_1}+e^{i\Delta \phi_2}+e^{i\Delta \phi_3}$.
The convention independent CP violation parameter $J_{CP}$
\cite{CECELIA} is given by:
\begin{equation}
J_{CP}=\frac{1}{27} {\rm Im}(e^{i(\Delta \phi_2-\Delta \phi_1)}
+e^{i(\Delta \phi_3-\Delta \phi_2)}+e^{i(\Delta \phi_1-\Delta \phi_3)})
\end{equation}
For example, if
$\Delta \phi_1 = 0^o$ and $\Delta \phi_2 = 60^o$
(and hence $\Delta \phi_3 = -60^o$) then $S=2$ and
$|p|= \sqrt{7}/3 \simeq 0.882$,
$|q|=|r|=1/3 \simeq 0.333$ and
$J_{CP}=1/(18\sqrt{3}) \simeq 0.032$.
We see no way to justify such a choice of phases however.

At this point, we return again
to the similarity we noted above,
between the operator eq.(1) and the
simple derivative operators representing
the ordinary kinetic terms in the Lagrangian.
Building on this observation, we now note that
a close analogy exists between the operator eq.(3)
and (the hermitian square of) a
full gauge-covariant kinetic operator.
The phases $\phi_i$ ($i=1$-3) play a role here
analogous to that of the gauge potential.
The freedom to change the absolute phases
using any (common) arbitrary diagonal matrix of phase factors,
is analogous to local gauge invariance.
A gauge-field configuration corresponding
to a constant field-strength (ie.\ a uniform field)
is of particular interest to us here, because
a uniform field is manifestly
translationally invariant.
We note that even in the case of a uniform field,
the inherent translational invariance cannot be explicit
in all of the components of the gauge potential at once,
after a choice of gauge has been made.
In the same way if we set:
\begin{equation}
\Delta \phi_2-\Delta \phi_1 = \Delta \phi_3-\Delta \phi_2 =
\Delta \phi_1-\Delta \phi_3
\end{equation}
corresponding to a uniform field
(in the discrete generation space),
then it must be that no generation is preferred,
even though the up-type and the down-type mass matrices
clearly cannot both be circulant.
As far as observables are concerned,
this last requirement eq.(7)
(together with the requirement
$\Delta \phi_1 + \Delta \phi_2 + \Delta \phi_3 = 0$, above)
completely specifies our ansatz
(eg.\
$\Delta \phi_1=0^o,\Delta \phi_2= \pm 120^o,\Delta \phi_3= \mp 120^o$),
up to the sign of $J_{CP}$.
The CKM matrix is flat in this case,
ie.\ all elements have equal modulus
$|p| = |q| = |r| = 1/\sqrt{3} \simeq 0.577$,
and $J_{CP}$ is extremal,
ie.\ $|J_{CP}|= 1/(6\sqrt{3}) \simeq 0.096$ \cite{CECELIA}.

If the above matrices are relevant at all,
they are relevant only at very high energy,
eg.\ unification (GUT) energies, and
have to be evolved down
to the electro-weak (EW) scale
in order to be compared with experiment.
The leading-order evolution equations \cite{BARGER}
for the quark Yukawa matrices in the Standard Model (SM)
can be written (neglecting the influence of the charged leptons):
\begin{eqnarray}
\dot{\alpha_u} & = &\frac{3}{2}\alpha_u^2
-\frac{3}{4}(\alpha_u\alpha_d+\alpha_d\alpha_u)
+3{\rm Tr}(\alpha_u+\alpha_d)\alpha_u-8\alpha_3\alpha_u
-\frac{9}{4}\alpha_2\alpha_u
-\frac{17}{20}\alpha_1\alpha_u \nonumber \\
\dot{\alpha_d} & = &\frac{3}{2}\alpha_d^2
-\frac{3}{4}(\alpha_u\alpha_d+\alpha_d\alpha_u)
+3{\rm Tr}(\alpha_u+\alpha_d)\alpha_d-8\alpha_3\alpha_d
-\frac{9}{4}\alpha_2\alpha_d
-\frac{5}{20}\alpha_1\alpha_d \\
\dot{\alpha_3} & = &-7\alpha_3^2 \nonumber \hspace{10mm}
\dot{\alpha_2}   =  -\frac{19}{6} \alpha_2^2 \hspace{10mm}
\dot{\alpha_1}   =  \frac{41}{10}\alpha_1^2
\end{eqnarray}
where Tr denotes the matrix trace,
the dot denotes differentiation with respect to
$T=(1/2\pi) \ln (E/E_0)$ and $E/E_0$ is the running energy scale,
expressed as a fraction of the starting energy.
The hermitian squares of the up-type and the down-type
Yukawa matrices are represented by $\alpha_u$
and $\alpha_d$ respectively, where
a factor of $1/4\pi$ has been incorporated in the definition
of $\alpha_u$ and $\alpha_d$ to simplify
the form of the evolution equations, in analogy with
the case of the gauge couplings.
The corresponding equations for the gauge couplings
($\alpha_i$, $i=$ 1-3) are included for completeness.

There has been much progress in understanding
the effects of evolution analytically \cite{GRZAD},
but for simplicity the results presented here
are based on a straightforward
numerical integration of eq.(8),
employing an appropriate (variable) stepsize.
Suitable starting values for the gauge couplings
are taken from the fits of Amaldi et al.\ \cite{AMALDI}.
For a given set of starting values for
the Yukawa couplings,
we calculate the quark mass spectrum and
the CKM matrix at the lower energy scale.
There is considerable freedom in
choosing starting values for the Yukawa couplings
consistent with the observed mass spectrum at low energies due
(in large part) to the well known
quasi-fixed-point \cite{ROSS}, implicit in the evolution equations,
which tends to focus the top Yukawa coupling towards
its fixed-point value at low energies,
independent of its starting value.
In spite of this, we find that
assuming {\em perturbative} starting values
for the individual Yukawa couplings
(ie.\ $\alpha_u,\alpha_d \simlt 1$),
chosen to reproduce the observed quark mass spectrum,
the predicted evolution is always
too slow to yield a realistic CKM matrix at low energies.
Evolving down over a reasonable range in $T$
(the GUT scale and the EW scale are
about five units apart in $T$)
the CKM matrix remains approximately flat;
that is to say,
all elements remain close to their starting value,
$|V_{ij}| \simeq 1/\sqrt{3} \simeq 0.577$,
to within deviations at the level of 20\% or less.

However, with recent experimental results
from LEP and from the Tevatron tending
to favour large values for the top mass \cite{TOP},
it is becoming increasingly clear that
the Yukawa couplings may very well be
{\em non-perturbative} at high energy.
Whilst we do not expect perturbative
evolution equations to be quantitatively valid
in a non-perturbative regime,
we have done what we can
to investigate this possibility,
by applying eq.(8) also in the case that
the Yukawa couplings assume non-perturbative values
(ie.\ $\alpha_u,\alpha_d \simgt 1$).
As one might expect, with larger starting values
for the Yukawa couplings, the evolution
proceeds more rapidly.
The observed quark mass spectrum at low energy,
can still be correctly reproduced,
thanks to the quasi-fixed-point.
We now find, however, that the CKM matrix,
although starting out absolutely flat,
rapidly develops a significant hierarchy which,
for suffiently large starting values for the Yukawa couplings,
is not-at-all unlike the familiar hierarchy \cite{WOLF}
of CKM amplitudes observed experimentally.
That said, we have not succeeded in finding
any one complete set of starting values which
reproduces the quark mass spectrum and the
CKM matrix simultaneously in every detail, and
in view of the strict inapplicabilty of eq.(8)
in the non-perturbative domain, neither should we expect to,
even in the case that our ansatz was perfectly correct.
Instead we give here a sample set
of starting values that can be seen to
reproduce most of the quark masses correctly,
together with the main features of the CKM matrix.
The input values for the (diagonalised) Yukawa couplings
at high energy are:
$\alpha_u=(6.0 \times 10^{-2}, 2.0 \times 10^{9}, 7.0 \times 10^{11})$,
$\alpha_d=(1.5 \times 10^{-1}, 5.0 \times 10^{0}, 4.5 \times 10^{1})$
leading to
$\alpha_u=(4.4 \times 10^{-11}, 8.3 \times 10^{-2}, 8.8 \times 10^{-2})$,
$\alpha_d=(2.8 \times 10^{-10}, 6.0 \times 10^{-8}, 6.8 \times 10^{-5})$
at the EW scale ($\Delta T =-5$).
The evolved CKM matrix is as follows
(only the moduli of the elements are given here;
phases are of course convention dependent):
\begin{equation}
V=\left(\matrix{
0.975 & 0.222 & 0.011 \cr
0.222 & 0.974 & 0.047 \cr
0.012 & 0.046 & 0.999 \cr
} \right)
\end{equation}
with $|J_{CP}| = 1.06 \times 10^{-4}$.
The result eq.(9) bears a
striking resemblance to the experimentally
observed CKM matrix and
suggests to us that it is evolution
(albeit non-perturbative and presently incalculable)
which is responsible for the observed hierarchy
in the CKM matrix at low energy.
Whilst results obtained
by applying perturbative equations in a non-perturbative domain
are unsatisfactory, in that they clearly cannot
be used to falsify any hypothesis at all,
we maintain that they do serve a useful purpose here
as an illustration of existing possibilities.
The problem of non-perturbative evolution
may not be forever intractable:
exact non-perturbative evolution equations
for coupling constants in pure gauge theories
have already been discussed in the literature \cite{EXACT}.
Certainly it cannot be said that
this ansatz is ruled out by experiment.
On the contrary, if the trends we see applying
leading-order perturbative evolution equations
are at all representative of the effects of complete
non-perturbative evolution, then all the indications
are that we are on the right track.

In conclusion,
in spite of the difficulties
we have emphasised,
we find the apparently natural emergence
of a CKM-like hierarchy entirely within
the SM framework very impressive.
The matrix operators we have proposed
come as close as one might
hope to generalising (to the discrete generation space)
the continuum gauge-covariant operators
already present in the SM Lagrangian.
One might even speculate that it is some analogue
of the pure-gauge kinetic term,
constructed from the relevant invariants \cite{CECELIA},
which (classically extremised)
accounts for the hierarchy of quark masses.
At the very least,
we believe that we have demonstrated that
this simple and appealing ansatz
merits further investigation.

\vspace{1cm}
\parindent 0mm
{\bf Acknowledgement}

It is a pleasure to thank R. G. Roberts for helpful discussions
and encouragement.

\end{document}